\documentclass[a4paper, amsfonts, amssymb, amsmath, reprint, showkeys, nofootinbib, twoside, superscriptaddress]{revtex4-2}
\usepackage{caption}
\usepackage{bm}
\usepackage{subcaption}
\usepackage[colorinlistoftodos, color=green!40, prependcaption]{todonotes}

\usepackage{amsthm}
\usepackage{mathtools}
\usepackage{dsfont}
\usepackage{physics}
\usepackage{xcolor}
\usepackage{graphicx}
\usepackage[left=23mm,right=13mm,top=35mm,columnsep=15pt]{geometry} 
\usepackage{adjustbox}
\usepackage{placeins}
\usepackage[T1]{fontenc}
\usepackage{lipsum}
\usepackage{csquotes}

\usepackage[pdftex, pdftitle={Article}, pdfauthor={Author}]{hyperref} 
\usepackage{hyperref}
\newcommand{\Berkeley}{Department of Physics, University of California, Berkeley, California 94720, USA}
\newcommand{\MIT}{Center for Theoretical Physics, Massachusetts Institute of Technology, Cambridge, MA 02139, USA}
\newcommand{\LBNL}{Materials Science Division, Lawrence Berkeley National Laboratory, Berkeley, CA 94720, USA}

\begin{document}
\title{A universal error correction crossover in semi-classical circuits}

\title{Universal information dynamics in semi-classical circuits}

\title{A universal crossover in quantum circuits governed by a proximate classical error correction transition}







\author{Anasuya Lyons}
\affiliation{\Berkeley}
\author{Soonwon Choi}
\affiliation{\MIT}
\author{Ehud Altman}
\affiliation{\Berkeley}
\affiliation{\LBNL}


\date{\today} 

\begin{abstract}
We formulate a semi-classical circuit model to clarify the role of quantum entanglement in the recently discovered encoding phase transitions in quantum circuits with measurements.  As a starting point we define a random circuit model with nearest neighbor classical gates interrupted by erasure errors. In analogy with the quantum setting, this system undergoes a purification transition at a critical error rate above which the classical information entropy in the output state vanishes. We show that this phase transition is in the directed percolation universality class, consistent with the fact that having zero entropy is an absorbing state of the dynamics; this classical circuit cannot generate entropy. Adding an arbitrarily small density of quantum gates in the presence of errors eliminates the transition by destroying the absorbing state: the quantum gates generate internal entanglement, which can be effectively converted to classical entropy by the errors. We describe the universal properties of this instability in an effective model of the semi-classical circuit. Our model highlights the crucial differences between information dynamics in classical and quantum circuits.
\end{abstract}


\maketitle

\section{Introduction}
Recent pioneering experiments manipulating a large number of qubits have brought questions on the resilience of quantum information in noisy circuits into sharp focus and motivated a theoretical effort to understand universal aspects of this dynamics~\cite{google,wu2021strong,zhu2022quantum,PhysRevLett.127.180502}. 
A case in point is the discovery of a measurement induced phase transition in circuits consisting of random unitary gates interrupted by occasional local measurements~\cite{skinner_measurement-induced_2019, li_quantum_2018, Chan2019, choi_quantum_2020, gullans_dynamical_2020,Noel2021, Koh2022}.
It was found that this system sustains large scale entanglement below a critical measurement rate marking the transition from volume-law to area-law scaling of the bipartite entanglement.  
The stability of the volume law phase is understood to be a result of the nonlocal encoding of information in the circuit, which protects quantum information from the disentangling effect of measurements \cite{choi_quantum_2020}. Thus, in the volume law phase
a finite fraction of the quantum information encoded in the initial state persists in the circuit for arbitrarily long times (in an infinitely wide circuit) in spite of the non-unitary element of the dynamics. 
\begin{figure}[t]
    \includegraphics[width=1\linewidth]{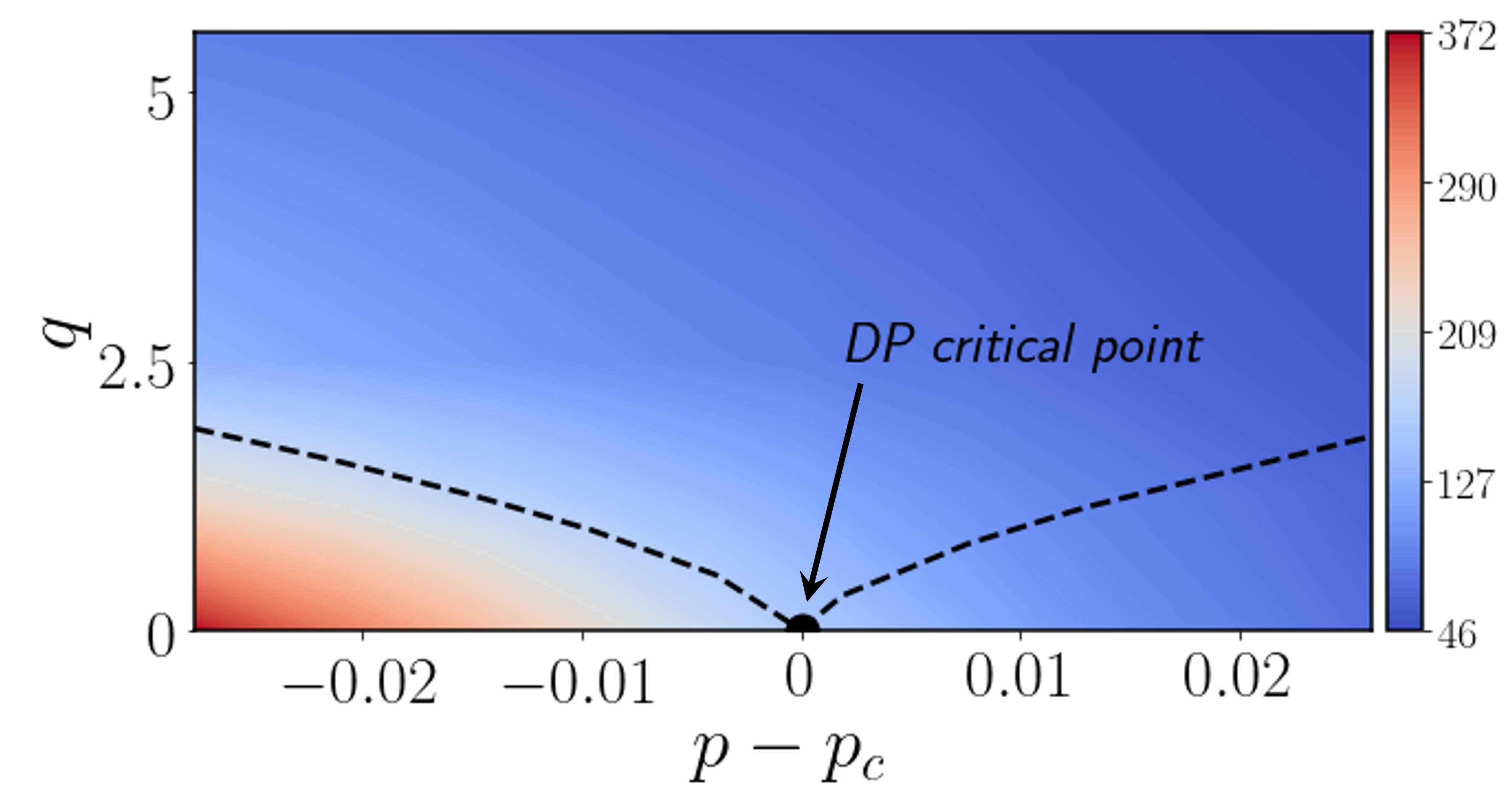}
    \caption{\textbf{Circuit phase diagram.} Phase diagram for our model; error rate $p$ on the horizontal and ``quantumness'' $q$ on the vertical. There is a sharp transition for $q=0$, at $p=p_c$ (classical critical point); dotted lines indicate the critical fan marking the quantum critical region. The background color shows the decay timescale for input:output mutual information ($N = 120$ qubits). In the coding phase, mutual information persists at long times, while in the non-coding phase, any initial encoded information is quickly lost.}
    \label{fig:fig1}
\end{figure}

In the measurement-induced transition,
the ``errors'' in unitary circuits are effected by active measurements performed by an observer, rather than decoherence or uncontrolled noise due to coupling to a passive environment. The quantum information encoded in the circuit is conditioned on the observer's knowledge of the measurement outcomes, and only by making use of this knowledge is it theoretically possible to retrieve the information. Thus, the protected encoding in this case relies on the unique nature of measurements in quantum mechanics --- measurements generally destroy off-diagonal quantum coherence while they simultaneously reveal some classical information encoded in the measurement outcomes.
To make a closer connection to realistic quantum dynamics it is important to understand the dynamics of encoded information under quantum circuits in the presence of uncontrolled errors.

\begin{figure*}[t]
    \includegraphics[width=1\linewidth]{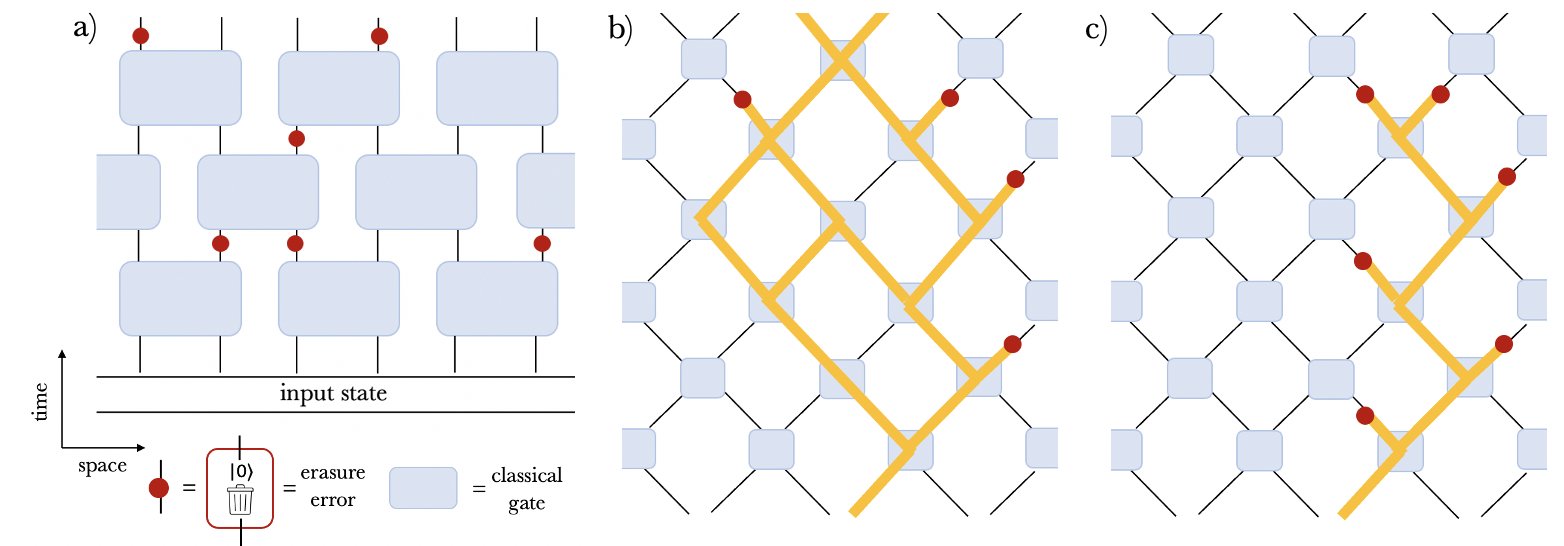}
    \caption{\textbf{Effective Directed Percolation Model.} Circuit diagram for our classical circuit model (a), mapping to a directed percolation problem (b), (c). Consider the dynamics of a single encoded bit (yellow bond). As it reaches a gate (lattice vertex), it stays the same, jumps one site, or branches into two encoded bits. Errors (red dots) break the bonds created by encoded bits. In b), the error rate is low enough that a percolation cluster reaches the end of the circuit. In c), the errors overwhelm the encoded bits and the cluster does not span the circuit. These demonstrate typical dynamics of an encoded bit in the encoding and non-coding phases, respectively.}
    \label{fig:fig2}
\end{figure*}

To address this question, we take one step back and characterize an encoding phase transition in random \emph{classical} circuits, where random measurements are replaced by erasures of bits, representing realistic errors.
After establishing that such classical circuits exhibit an information encoding phase transition, we then introduce quantum effects perturbatively to determine their impact on the classical encoding transition. The vicinity of the classical critical point allows a controlled description of information dynamics in the quantum circuit (see \ref{fig:fig1}).

The classical circuit we consider, shown in Fig. \ref{fig:fig2}(a), consists of random two bit gates operating on nearest neighbors in a chain of classical bits. An erasure error that resets the bit to a value $0$ can occur with probability $p$ on each site following application of a logic gate. 
Without the errors, the circuit implements a reversible unitary transformation on the input bit-string. Thus all the information encoded initially is also present in the scrambled output string, regardless of the circuit depth.
In the presence of errors the information content in the output state is degraded, but vanishes only above a critical error rate. 
We present theoretical arguments and numerical evidence that this classical encoding transition is in the directed percolation universality class. The state with vanishing information entropy plays the role of an inescapable absorbing state. If the information on the input state is gone it cannot be resurrected. 

The classical model can also be viewed as a quantum circuit with unitary gates that do not entangle computational basis states (aka bit-strings)~\cite{iaconis_measurement-induced_2020}. We introduce entanglement growth in a controlled way by adding a small density $q$ of Hadamard gates. Once the system is entangled,  erasure errors can generate entropy even from a pure state. Therefore the zero entropy state is no longer an absorbing state in the presence of the Hadamard gates. 
We demonstrate numerically that the quantum perturbation $q$ has the scaling properties of an additive noise term near the directed percolation critical point. Thus the quantum system shows an error induced universal crossover governed by the critical point located on the classical axis $q=0$. 

\section{Random classical circuits}

We consider a classical model of 2-bit gates operating on a string of $N$ bits, ${\mathbf x} = x_1x_2, \dots x_N $. The gates are arranged in a brick-wall structure in space-time as illustrated in Fig. \ref{fig:fig2}(a). Each gate implements a permutation of the four possible 2-bit input states $\{00,01,10,11\}$, represented by a $4\times 4$ unitary matrix $g$. Gates are chosen at random from independent uniform distributions over the permutation group $S_4$.
In addition to reversible gates, bit-strings are subject to erasure errors occurring at every position in each time step with probability $p$.
The bit at an error site is reset to the $0$ state. 
We are interested in the dynamics and protection of information in such circuits. Specifically, if the input state encodes a certain amount of information, what fraction of this information survives after $t$ time steps? We show that in the thermodynamic limit of infinitely wide arrays, the information surviving at long times vanishes at a critical error rate $p_c$ and that the transition is in the directed percolation universality class. For $p<p_c$ a finite fraction of the information is protected from errors. 

It is instructive to first consider a simple initial state that encodes one bit of information: ${\mathbf x}_{in}=(0,\ldots,0,x_j,0\ldots 0)$. The encoded site $j$ hosts a random variable taking a value $0$ or $1$ with equal probabilities, while all other sites take a definite value. The reversible gates spread the encoded logical bits over an increasing number of physical sites, in effect duplicating the information, but not changing the information content. The encoding state at a given time is represented by a joint probability distribution $P(x_1,\ldots,x_N)$ and the information content is quantified by the associated Shannon entropy. 

The reversible time evolution, effected by the gates, gives rise to a growing cluster of sites hosting indefinite bits, as depicted in Fig. \ref{fig:fig2}(b,c). 
Erasure errors reset the bit in the physical error site, giving it a definite value 0, thereby removing part of the duplicated information and stunting the growth of the ``information cluster''. 
As long as indefinite bits remain, however, a fraction of the information survives with them. 

It is natural to expect an error threshold, above which the errors prevent information clusters from growing to infinite size, as seen in Fig. \ref{fig:fig2}(b). Once all encoded bits are destroyed all the encoded information is permanently erased.
The growth of the information cluster follows the general scheme of a directed percolation (DP) process or an equivalent population dynamics model \cite{broadbent_hammersley_1957, Harris_DP, 2000}. The indefinite bits can be viewed as a population of bacteria that move and multiply through the action of the reversible gates, and are killed by erasure errors. 

\subsection{Diffusion-Reaction Model}

While this illustration considers only one bit of information initially encoded at site $i$, it is straight forward to generalize to translationally invariant initial states with a finite density of encoded information.
In fact, the dynamics of the purity of the probability distribution, averaged over random instances of classical circuits, can be mapped exactly to a stochastic diffusion-reaction model in the DP class.
Using this model, it can be established that the information encoding/non-encoding phase transition, defined as follows, belongs to the directed-percolation universality class.
More specifically, we consider $N$ bits initially in a uniform probability distribution, $p(x) = 1/2^N$ with $x\in \{0,1\}^{\otimes N}$, undergoing random classical circuits in the presence of erasure errors.
In order to identify a phase transition we evaluate the dynamics of the collision probability $Q\equiv \sum_x p^2(x)$ of bitstring distributions at the output, averaged over the choice of random classical gates in the presence of erasure errors occurring at predetermined locations.
Generalizing our analysis to a random location of  erasure errors should be straight-forward.
We ask if the average of $Q$ quickly approaches $1$, where all information in the circuit is lost, or it remains exponentially small in $\sim N$ for a long time, where $\sim -\log_2(Q)$ bits of information remain protected in the circuit.
The average $Q$ is directly related to the annealed averaged Renyi-2 entropy
\begin{align}
     \overline Q = \overline{e^{-S_2} }.
\end{align}
Our main result will be that the dynamics of $\overline Q$ as a function of circuit depth is governed by a diffusion-reaction model whose phase transition belongs to the directed percolation universality class.

\begin{figure*}[t!]
\includegraphics[width=1\textwidth]{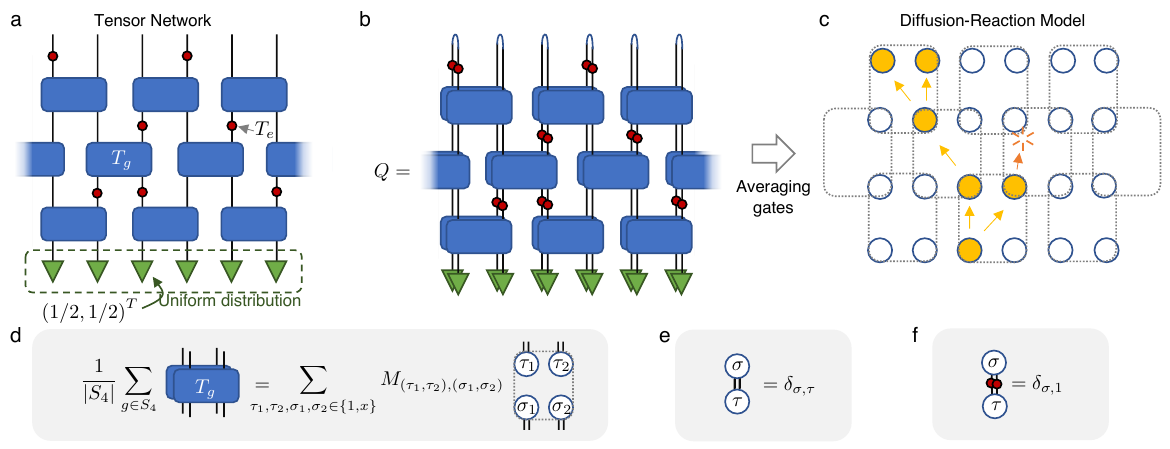}
	\caption{(a) Tensor-network representation of random classical circuits with erasure errors.
	(b) The duplicated tensor network diagram allows evaluating the collision probability $Q$.
	(c) Averaging over random classical gates leads to a diffusion reaction model.
	(d) Averaging a single random classical gate gives rise to the summation over classical configurations of $\sigma$ and $\tau$ variables $\in \{1, x\}$ weighted by $M$. In the language of the diffusion-reaction model, $\sigma=1$ or $\sigma = x$ corresponds to the absence or the presence of a particle at the site. The weight $M$ can be considered as a Markov map describing the diffusion and reaction process of the particles.
	(e,f) Useful identities to derive the diffusion-reaction model.}
	\label{fig:TN_gate_avg}
\end{figure*}

Our strategy is the following.
First, we consider the tensor-network representation of random classical circuit evolution.
Tensors in this section are associated with probability distributions rather than quantum wavefunctions.
In this representation, we encode $p(x)$ using a network of tensors with total $N$ open legs of dimension 2 (running over $0$ and $1$).
The open legs enumerate over $2^N$ bitstrings $\{x\}$, and the contraction of the network against a given assignment $x$ for the open legs evaluates $p(x)$.
At initial time (zero circuit depth), the uniform probability distribution is factorizable, hence can be represented by $N$ separate tensors. 
A classical gate operation $g$ corresponds to a $4\times 4$ matrix $T_g$ where
the indices $\alpha$ and $\beta$ run over four possible bitstrings $\{00, 01, 10, 11\}$
and
$\left(T_g\right)_{\alpha \beta} = 1$ if $g(\beta) = \alpha$ and $0$ otherwise.
Since we only consider reversible gates, $T_g$ is equivalent to the linear representation of the symmetry group $S_4$.
The erasure error acting on a single bit can be described a $2 \times 2$ matrix
\begin{align}
    T_e = \left(
    \begin{array}{cc}
    1 & 1\\
    0 & 0
    \end{array}
    \right).
\end{align}
$T_e$ maps any local input bit values, i.e., either `0' represented by $(1,0)^T$ or `1' by $(0,1)^T$, to `0' ($(1,0)^T$).
As explained in the main text, this error is equivalent to (i) summing over two possible values of a bit, followed by (ii) inserting a fresh bit set to the constant `0'. 
The diagrammatic representation is shown in Fig.~\ref{fig:TN_gate_avg}.

Second, we consider two copies of identical classical circuits and erasure errors represented in tensor network diagrams.
By properly contracting two networks, we obtain a joint tensor network whose contraction evaluates the collision probability $Q$ for a particular instance of random circuit and erasure errors.
The exact contraction of such tensor network amounts to simulating the dynamics of the classical circuits.

Finally, we perform averaging over all possible choices of random permutations $g\in S_4$.
This can be done without evaluating the full tensor network diagram for individual circuit realizations.
In particular, we utilize the following identity
\begin{widetext}
\begin{equation}
\begin{aligned}
\label{eqn:gate_avg}
    \frac{1}{|S_4|} \sum_{g\in S_4} T_g \otimes T_g  
= \sum_{\sigma_1,\sigma_2,\tau_1, \tau_2 \in \{1, x\}}
M_{(\tau_1,\tau_2), (\sigma_1,\sigma_2)} \hat{\tau}_1 \otimes 
\hat{\tau}_2 \otimes
\hat{\sigma}_1 \otimes
\hat{\sigma}_2
\end{aligned}
\end{equation}
\end{widetext}
where $|S_4|=24$ is the number of elements in the permutation group $S_4$;
$\hat{\tau}$ and $\hat{\sigma}$ are $2\times 2$ matrices  defined as \begin{equation}
\begin{aligned}
    \hat{\sigma} =
    \left\{
    \begin{array}{cc}
    \mathds{1} & \textrm{if } \sigma = 1\\
    X & \textrm{if } \sigma = x
    \end{array}
    \right.
\;\;
\&
\;\;
    \hat{\tau} =
    \left\{
    \begin{array}{cc}
    \frac{1}{2}\mathds{1} & \textrm{if } \tau =  1\\
    \frac{1}{2} X & \textrm{if } \tau = x
    \end{array}
    \right.
\end{aligned}
\end{equation}
with the Pauli opertor $X := (0, 1;1,0)$; and
the weight matrix 
\begin{align}
    M = \left(
    \begin{array}{cccc}
        1 & 0 & 0 & 0\\
        0 & 1/3 & 1/3 & 1/3\\
        0 & 1/3 & 1/3 & 1/3\\
        0 & 1/3 & 1/3 & 1/3
    \end{array}
    \right)
\end{align}
in the order $(1,1),(1,x),(x,1),(x,x)$.
The relation in Eq.~\eqref{eqn:gate_avg} can be checked by explicit calculation.
The diagrammatic representation of Eq.~\eqref{eqn:gate_avg} is shown in Fig.~\ref{fig:TN_gate_avg}(d).
Repeatedly applying Eq.~\eqref{eqn:gate_avg} for every gate, we obtain an expression that resembles a partition sum over variables $\sigma, \tau \in \{1,x\}$.

The partition sum can be simplified.
The contraction of $\hat{\sigma}$ and $\hat{\tau}$ tensors arising from two successive gates evaluates to the kronecker delta function $\delta_{\sigma, \tau}$.
Also, the erasure error in the classical circuit leads to a kronecker delta function $\delta_{\sigma,1}$, which is equivalent to forcing the corresponding $\sigma$ variable to take the value $1$ independent of the value of the contracted $\tau$ variable.
Using these kronecker delta functions, one can eliminate all $\tau$ variables from the partition sum except those at the top boundary. The $\tau$-tensors at the top boundary shall be traced out as we will describe below.
The final expression for the bulk of our diagram only involves the partition sum over all possible configurations of $\sigma$ variables weighted by products of $M$, each depending on four $\sigma$ variables at its corners [Fig.~\ref{fig:TN_gate_avg}(c)].

The partition sum can be interpreted as a path integral of a diffusion-reaction model.
More specifically, we note that the row and column sums of the weight matrix $M$ are unity, hence $M$ describes a Markov process. 
One can interpret $\sigma_i = 1$ or $\sigma_i = x$ as a particle being absent or present at the site $i$, respectively. 
Then, the weight matrix $M$ describes a local Markov process: (i) if two neighboring sites are both empty, e.g. $(\sigma_i, \sigma_{i+1}) = (1,1)$, then they remain in the same configuration, and
(ii) if at least one of the two sites is occupied by a particle, e.g. $(\sigma_i, \sigma_{i+1}) \in\{ (1,x), (x,1), (x,x)\}$, the two sites will take one of the three configurations with probability $1/3$ after the Markov process.
Here (ii) describes both diffusion and reaction processes of particles.
The presence of erasure error in the classical circuit effects the spontaneous loss of particles.
This model has an absorbing state, where every site is empty.

After time evolution, $\tau$ tensors are contracted with top boundary tensors that evaluate the collision probability in the classical circuit. 
In terms of the diffusion-reaction model, such contraction is equivalent to taking the trace of all $\tau$ tensors at the top in Fig.~\ref{fig:TN_gate_avg}.
\begin{align}
    \overline{Q} = \sum_{\{ \tau_j \}}
    p(\{\tau_j\})
    \prod_j  \textrm{tr}\{\tau_j\},
\end{align}
where the summation is over all possible configurations of $\tau$ variables at the top boundary and $p(\{\tau_j\})$ is the probability distribution of each configuration, obtained from the Markov process in the bulk.
Since the Pauli $X$ operator is traceless, the only non-vanishing contribution to $\overline{Q}$ arises when every $\tau_j = 1$.
Therefore, 
\begin{align}
    \overline{Q} = \textrm{Pr}(\tau_j =1 \textrm{ at every site $j$}).  
\end{align}
Apparently, this quantity undergoes a phase transition in the diffusion-reaction model.
When the particle loss rate is sufficiently large, the system reaches the absorbing state with high probability, and $\overline{Q} \approx 1$ (non-coding phase).
However, if the loss rate is sufficiently small, the particles proliferate and $\overline{Q}$ remains small.
Note that, due to the initial configuration, i.e. every $\sigma$ variables are $1$ and $x$ with equal probability, the system starts in the absorbing state with the probability $1/2^N$. Therefore, $\overline{Q}$ is at least $1/2^N$ at all time, which is consistent with the the fact that the collision probability cannot be smaller than $1/2^N$.

\subsection{Numerical Results}

We validate the prediction for an encoding phase transition in the DP universality class by direct simulation of the circuit model. As the reversible gate operations form a sub-set of the Clifford group (with only $\{ \mathbb{I}, \sigma^z \}$ stabilizers), we can use the stabilizer formalism \cite{aaronson_improved_2004} to efficiently simulate the circuit. Generalization to a quantum circuit model is straightforward using this approach. 
For now, we focus on the classical circuit dynamics. We initialize the system in a maximal entropy state (uniform distribution over all $2^N$ bitstrings).
The entropy of the state at each subsequent time step is computed using standard techniques \cite{audenaert_entanglement_2005, shi2020}.

\begin{figure}
    \centering
    \includegraphics[width=1\linewidth]{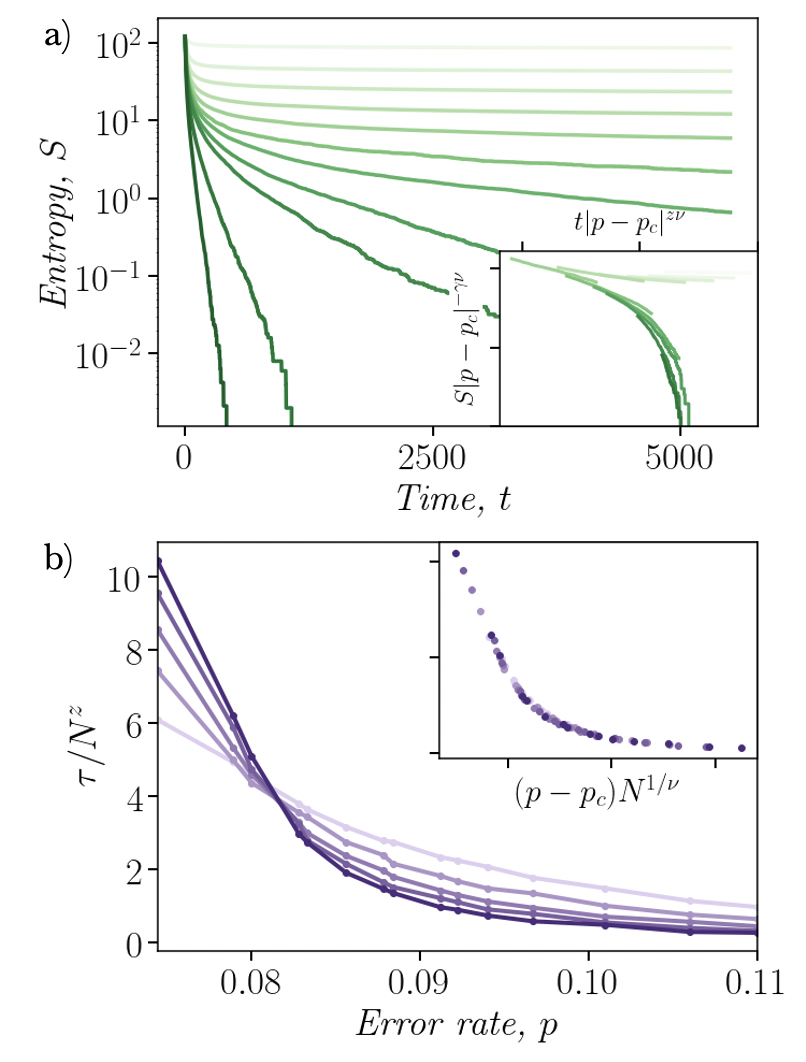}  
    \caption{\textbf{Classical circuit dynamics.} a) Example system entropy versus time curves with $q = 0$, $N = 120$. Darker color corresponds to higher $p$; for small $p$, the entropy decays to a finite value, while for larger $p$, it decays to zero. The inset shows the scaling behavior of the entropy vs. time curves, with different scaling laws for $p < p_c$ and $p > p_c$. We find $\gamma = 0.75$, using $z$, $\nu$ extracted from the scaling in (b). b) Scaling for the decay timescale $\tau$ at different error rates and system sizes ($N = 40$ to $N = 120$ in increments of $20$, light to dark). Rescaling the y-axis yields dynamical exponent $z = 1.51$ and critical error rate $p_c = 0.081$. X-axis rescaling (inset) recovers exponent $\nu = 1.1$. }
    \label{fig:fig3}
\end{figure}

Figure \ref{fig:fig3}(a) shows the decay of the entropy (information) with time for different values of the error rate.  
We see these curves follow the expected behavior: for rates $p < p_c$, the entropy decays only slightly to some finite amount, for $p > p_c$ the entropy decays to zero over a finite time scale $\tau$. To determine $\tau$ we let the system evolve beyond a time $t_0$, at which transient effects become negligible, and define $\tau$ to be the time it takes the entropy to decay to $15\%$ of its value at $t_0$. The transition is signaled by a divergence of $\tau$, which is cut off at $\tau\sim N^z$ in a system with $N$ bits. Fig. \ref{fig:fig3}(b) shows a good crossing of the rescaled decay time $\tau(p)/N^z$, allowing us to extract the critical error rate $p_c=0.081$ and dynamical exponent $z=1.51$. The inset shows good data collapse using the scaling ansatz $\tau = N^z \mathcal{F}((p-p_c) N^{\frac{1}{\nu}})$ with the extracted $z$, $p_c$, and correlation length exponent $\nu=1.1$.
These values are consistent, within errors, with the accepted exponents of the directed percolation universality class $z=1.58$ and $\nu = 1.09$~\cite{2000}. 

\begin{figure}
    \includegraphics[width=1\linewidth]{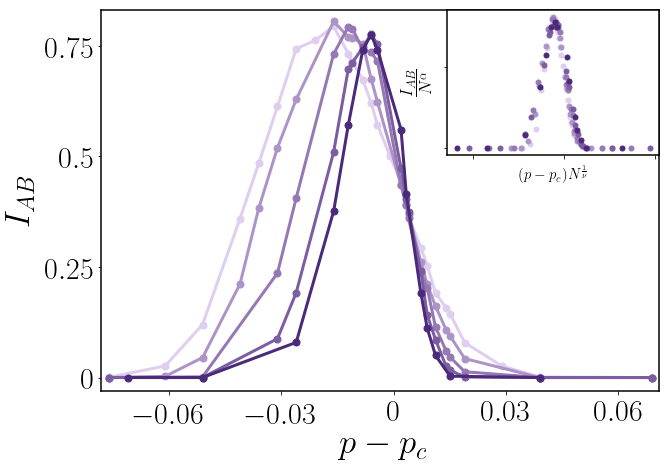}  
    \caption{Antipodal mutual information as a function of error rate for different system sizes $N$, from $N = 60$ to $N = 200$ going from left to right. We considered segments of length $\frac{N}{4}$, and evaluated mutual information at $t = N^{z}$, where $z = 1.51$ is our previously extracted dynamical exponent. The inset shows the scaling collapse with critical exponent $\nu = 1.2$ which is consistent with the directed percolation universality. Here, $\alpha = 0.28$. Note that the apparent crossover in the unscaled data is not relevant.}
    \label{fig:apdx_fig1}
\end{figure}

As an additional check on our results, we studied the dynamics of antipodal mutual information in our circuit. Considering the mutual information between antipodal segments of length $N/4$ of our system as a function of error rate $p$ (see figure \ref{fig:apdx_fig1}), we found the expected peak around our critical value $p_c = 0.081$--- the actual peaks are shifted slightly by finite size effects, but they converge to $p_c$ in the thermodynamic limit. The critical exponent $\nu = 1.2$ is consistent with our previous results. 

The qualitative behavior of the mutual information also matches with our understanding of information dynamics in this model. For error rates above the critical point, the errors overwhelm the unitary gates and very little information is spread over long distances, leading to almost no mutual information between antipodal regions. For error rates below the critical point, unitary gates are able to scramble information very efficiently with little interference from errors; this also leads to low mutual information between any two regions. In the thermodynamic limit, there should only be nonzero mutual information at the critical point.

\section{Semi-classical circuits}
Having characterized an encoding phase transition in a classical model, our next task is to extend the analysis to quantum circuits that can approach the classical limit continuously.
This extension involves two important elements. First, we must reinterpret the classical circuit elements as operating on quantum bits.
Second, we must add gates such that the circuit can build up quantum entanglement. Here we restrict these to Clifford gates to facilitate efficient computation.

Classical gates were defined by their action on strings of classical Pauli operators $\mathbb{I}$ and $\sigma^z$. To embed the classical gates in a quantum circuit we should also specify how they act on strings containing $\sigma^x$ and $\sigma^y$.  The ``classicality'' constraint that the space of $\sigma^z$ strings remains closed under the action of the gates is not sufficient to fully specify the extension. For example, classical gates implementing the identity permutation of two bits can be extended not only to the identity unitary, but also to any unitary that rotates about the $\sigma^z$ axes of the two qubits. This freedom allows us to choose the extension of the  
classical gates in such a way that the space of Pauli strings involving only $\sigma^x$ is also closed under their action. 
This condition together with the requirement that Clifford gates preserve the commutation relations between the stabilizers, completely determines the action of the gates on the full Hilbert space. 

The errors also need to be interpreted within a quantum framework. The quantum channel corresponding to erasure errors can be implemented by operating a swap between the affected qubit and an environment ancilla qubit initialized to the state $|0\rangle$, then tracing over the ancilla. When this quantum channel operates on a non-entangled (i.e. classical) pure state the output is a pure classical state, which coincides with the effect of a classical erasure error. On the other hand, if the affected qubit is entangled with other system qubits, the entanglement is transferred to the environment and performing the trace will add to the entropy of the system. Thus, in a quantum model with entangled states propagating in the circuit, the entropy does not quantify the amount of previously encoded information and can be generated by an error, even if the error acts on a pure state of the system.  
But to have entanglement in the circuit in the first place, the ``classical gates'' discussed so far must be supplemented by gates that create coherence between computational basis states. Within the stabilizer picture, this translates to mixing between $\sigma^z$ and $\sigma^x$ strings, which can be achieved by introducing single-qubit Hadamard gates. The combined action of the Hadamards and the two-qubit classical gates can now produce entanglement inside the system. By tuning the probability $q$ of applying Hadamard gates, we control the rate of entanglement generation;  $q = 0$ will return us to the classical circuit model. 

Having a nonvanishing $q$ has a dramatic impact on the information dynamics in the circuit. Because acting on an entangled state with an erasure error can produce entropy even if the the error acts on a pure state, the zero entropy state no longer provides an absorbing state needed to protect the directed percolation phase transition. Within the effective theory of the transition we expect $q>0$ to produce an additive noise term, which is a relevant perturbation to the directed percolation fixed point. This perturbation leads to broadening of the transition to a universal crossover, which is governed by the underlying phase transition at $q=0$. 

In particular for $p=p_c$ the strength $q$ of the quantum term generates the only scale for the saturation timescale and value of the entropy. This suggests the scaling form for the decay of the entropy at $p=p_c$:
$S(t,q) \sim q^{\gamma/\eta}\mathcal{F}(tq^{z/\eta})$. Here $\eta$ is the scaling dimension of the perturbation $q$ and $\gamma$ the scaling dimension of the entropy at the directed percolation critical point. If $q$ behaves as an additive noise perturbation to the DP critical point, then we expect $\eta = 2.34$ and $\gamma = 0.75$ \cite{2000}. Note that we already found an entropy scaling consistent with $\gamma = 0.75$ when rescaling the classical entropy versus time curves (see inset of figure \ref{fig:fig3}a).


\begin{figure}
    \includegraphics[width=1\linewidth]{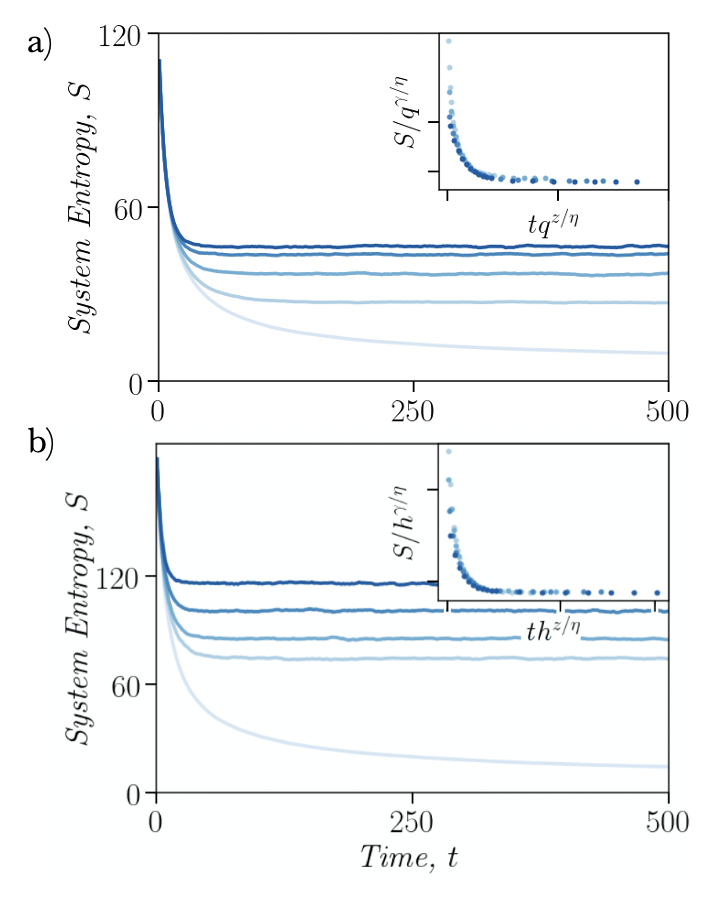}  
    \caption{\textbf{Addition of quantum gates and classical noise.} a) Entropy versus time curves for $N=120$, different $q$ ($q=0$ to $q=0.05$, bottom to top). We find $\eta = 2.34$ and $\gamma = 0.75$. Exponents $\eta$ and $\gamma$ agree with the known directed percolation exponents \cite{2000}. (In the cited reference, the exponents $\eta$ and $\gamma$ are given by $\sigma/\nu_{\perp}$ and $1 - \beta/\nu_{\perp}$ respectively.) b) Entropy vs. time curves for $p = 0.81$, $N = 200$, and different $h$ ($h=0$ to $h=0.05$ from bottom to top). As when adding Hadamard gates, $h>0$ leads to a finite saturation value of entropy. The scaling collapse is shown in the inset, using the same scaling relation as for the semi-classical circuits. Our exponents agree with our previous results and with the expected directed percolation values: $z = 1.58$, $\gamma = 0.75$, $\eta = 2.34$.}
    \label{fig:fig4}
\end{figure}

We verify these theoretical expectations by direct simulation of the stabilizer circuits.
Figure \ref{fig:fig4} shows the evolution of the system entropy $S$ for different values of $q$, at the critical error rate $p_c = 0.081$. When $q > 0$, $S$ decays to a finite equilibrium value rather than decaying to zero. A good scaling collapse is obtained using the exponents $\eta = 2.34$ and $\gamma = 0.75$, consistent with the expected directed percolation exponents (see inset of Fig. \ref{fig:fig4}). This confirms that the quantum effects leading to entanglement act as a relevant additive noise perturbation at the directed percolation critical point. 

Additionally, it can be shown that a classical additive noise process acting on a classical circuit has the same scaling behavior: we also considered adding a simple,  completely classical noise process to the circuit. We implemented this noise process by allowing bits with a defined value to become indeterminate with some probability $h$ at each time step--- we can now introduce encoded bits with junk entropy. Now, in addition to the probability $p$ that a bit will be replaced with a $0$, we have the probability $h$ that the bit will be erased and replaced with a junk encoded bit. Entropy can now be generated inside our circuit; we expect the absorbing state to be destroyed just as with the addition of Hadamards.

Allowing both erasure errors and this noise process, we study the behavior of the system entropy at the critical point $p_c = 0.081$ as a function of time and of noise strength $h$ (see figure \ref{fig:fig4}b). As in the quantum case, when $h>0$, there remains finite entropy in the system at long times even for $p>p_c$. The critical exponents governing the scaling behavior for $h$ and $q$ match as well (see the inset of figure \ref{fig:fig4}b). 

Even though the absorbing state in the entropy disappears for $q>0$, we might ask if another quantity, like the initial:final mutual information still retains one. In the classical model, measuring the system entropy was equivalent to measuring the mutual information. This equivalence no longer holds for $q>0$, meaning the mutual information might capture the transition while the entropy does not. However, while the mutual information does still have an absorbing state for $q>0$, any $p>0$ will drive the system into this absorbing state. This is because monogamy of entanglement, coupled with the infinite size of the bath, ensures that for times larger than order $N$ bath-system correlations dominate the reference-system correlations (the reference encodes the initial state). Thus, the memory of the initial state is always lost for $q>0$; the universal crossover discussed in this work is a crossover in the timescale at which this happens (see Fig. \ref{fig:fig1}).

\section{Discussion}
We have 
introduced a family of semiclassical quantum circuits, which allowed us to controllably add entanglement and study it’s impact on the dynamics of information encoding.
In the strictly classical limit the circuits undergo an encoding transition in the directed percolation class at a critical error threshold. 
Adding a small density of Hadamard gates, which allow buildup of entanglement in the circuit, behaves as an additive noise perturbation on the DP critical point. Thus quantum effects eliminate the error threshold and broaden the transition to a universal crossover. In the crossover regime, decay of the encoded information is controlled by the critical exponents of the underlying  DP transitions.

An important open question is whether quantum circuits with more structure can protect quantum information for arbitrary long times. The proximity to the classical critical point can help to identify what structures can counter act the relevant perturbation and resharpen the transition. 

The semi-classical circuits introduced here offer a new approach for studying the crossover from quantum to classical dynamics, while using powerful tools--- such as mapping to statistical mechanics models--- developed for random circuits. For example, such studies may shed light on the relation between classical and quantum chaos.

\noindent{\it Note Added:} While this paper was being written, a preprint appeared showing a directed percolation transition in chaotic classical circuits~\cite{willsher2022measurement}. The effect of quantizing the classical circuit has not been considered previously.


\begin{acknowledgements}

We thank Samuel Garratt, Yimu Bao, and Zack Weinstein for helpful discussions. We acknowledge support from the NSF QLCI program
through grant number OMA-2016245 (E.A.), and the Berkeley Physics Undergraduate Research Scholarship (A.L.). This research was done using services provided by the OSG Consortium \cite{osg07, osg09, osgDOI}, which is supported by the National Science Foundation awards \#2030508 and \#1836650.

\end{acknowledgements}



\bibliography{literature.bib} 
\bibliographystyle{unsrt} 



\end{document}